\documentclass[final,3p,times,twocolumn]{elsarticle}
 \biboptions{comma,sort&compress}
\usepackage{here}
 \usepackage{graphicx}
  \usepackage{epsfig}
\usepackage[english]{babel}
\usepackage{fancyhdr}
\usepackage{amsmath}
\usepackage{amssymb}
\usepackage{amsfonts}
\usepackage{psfrag}
\usepackage[applemac]{inputenc}
\usepackage{graphicx,wrapfig}
\usepackage[bf,footnotesize]{caption2}
\usepackage{pstricks}
\usepackage{multicol}
\usepackage{enumerate}
\usepackage{color}
\usepackage{array}
\usepackage{url}
\usepackage[title,titletoc,toc]{appendix}
\usepackage{float}
 \usepackage{hyperref}
 \usepackage{graphicx}
 \usepackage{subfigure}

\def\nin{\noindent}
\def\beq{\begin{equation}}
\def\eeq{\end{equation}}
\def\bea{\begin{eqnarray}}
\def\eea{\end{eqnarray}}

\newcommand{\bag}{\begin{align}}
\newcommand{\eag}{\end{align}}

\newcommand{\eq}[1]{Eq.~(\ref{#1})}
\newcommand{\fig}[1]{Fig.~\ref{#1}}

\newcommand{\nn}{\nonumber}

\newcommand{\al}{\alpha}

\journal{Nuc. Phys. (Proc. Suppl.)}

\begin{document}

\begin{frontmatter}



\title{The muonic hydrogen Lamb shift and the proton radius}

 \author{Clara Peset\corref{cor1}}
  \address{Grup de F\'{\i}sica Te\`orica and IFAE, Universitat Aut\`onoma de Barcelona, 08193 Bellaterra, Barcelona}
\cortext[cor1]{Speaker}
\ead{peset@ifae.es}



\begin{abstract}
\noindent
We obtain a model independent expression for the muonic hydrogen Lamb shift up to $\mathcal O (m_\mu \alpha^6, m_\mu \alpha^5 \frac{m_\mu^2}{m_\rho^2})$. The hadronic effects are controlled by the chiral theory, which allows for their model independent
determination. We give their complete expression including the pion and Delta particles. Out of this analysis and the experimental measurement of the muonic hydrogen Lamb shift we determine the electromagnetic proton radius: $r_p$=0.8412(15) fm. This number is at 6.8$\sigma$ variance with respect to the CODATA value. The parametric control of the uncertainties allows us to obtain a model independent determination of the error, which is dominated by hadronic effects.

\end{abstract}


\end{frontmatter}


\section{Introduction}
\nin
The recent measurement of the muonic hydrogen ($\mu p$) Lamb shift, $E(2P_{3/2})-E(2S_{1/2})$ \cite{Pohl:2010zza,Antognini:1900ns},
\begin{eqnarray}
\label{DeltaEexp}
&&\Delta E^{\rm exp}_L=202.3706(23)\,\mathrm{meV}
\end{eqnarray}
and the associated determination of the electromagnetic proton radius:
$r_p= 0.84087(39)$ fm has led to a lot of controversy. The reason is that this number is 4-7$\sigma$ away from previous determinations of this quantity coming from hydrogen and electron-proton ($ep$) scattering \cite{Mohr:2012tt,Lorenz:2014vha}. 

\nin
In order to asses the significance of this discrepancy it is of fundamental importance to perform the computation of this quantity
in a model independent way. In this respect, the use of effective field theories (EFTs) is specially useful. They help organizing the computation by providing with power counting rules that asses the importance of the different contributions. This becomes increasingly necessary as higher order effects are included. Even more important, these power counting rules allow to parametrically control the size of the higher order non-computed terms and, thus, give an estimate of the error. 

\nin
The EFT approach is specially convenient in the case of bound states where there are different, well separated scales, namely, the hard scale or reduced mass ($m_r$), the soft scale or typical momentum ($m_r v \sim m_r \alpha$) and the ultrasoft scale or typical binding energy ($m_r v^2 \sim m_r \alpha^2$).

\nin
In the case of $\mu p$ we need to deal with several scales:
\beq
\nn
m_{p} \sim m_{\rho},
\quad
m_{\mu} \sim m_{\pi} \sim m_r\equiv\frac{m_{\mu}m_p}{m_p+m_{\mu}},
\quad
m_r\alpha \sim m_e.
\eeq
from which we obtain the main expansion parameters by considering ratios of them
\beq
\label{ratio1}
\frac{m_{\pi}}{m_p} \sim \frac{m_{\mu}}{m_p} \approx \frac{1}{9}
\,
\frac{m_e}{m_r} \sim \frac{m_r\alpha}{m_r}\sim \frac{m_r\alpha^2}{m_r\alpha}\sim \alpha \approx \frac{1}{137}
\,.
\eeq
These, together with the counting rules given by the EFT provide the necessary tools to perform the full analysis of the Lamb shift in $\mu p$ up to leading-log $\mathcal O (m_\mu \alpha^6)$ terms and leading $\mathcal O (m_\mu \alpha^5 \frac{m_\mu^2}{m_\rho^2})$ hadronic effects.

\nin
In our approach we combine the use of Heavy Baryon Effective Theory (HBET) \cite{Jenkins:1990jv,Bernard:1992qa}, Non-Relativistic QED (NRQED) \cite{Caswell:1985ui} and, specially, potential NRQED (pNRQED) \cite{Pineda:1997bj,Pineda:1997ie,Pineda:1998kn}. Partial results following this approach can be found in \cite{Pineda:2002as,Pineda:2004mx,Nevado:2007dd} (see \cite{Pineda:2011xp} for a review). In Ref. \cite{Peset:2014yha} we computed the $n=2$ Lamb shift with accuracy $\mathcal O (m_\mu \alpha^6, m_\mu \alpha^5 \frac{m_\mu^2}{m_\rho^2})$. A more detailed account of the hadronic part can be found in \cite{Peset:2014jxa}. These proceedings are based on the work carried out in Refs. \cite{Peset:2014yha,Peset:2014jxa}.

\section{Lamb shift and extraction of the proton radius}
\nin
All contributions to the Lamb shift up to the order of our interest are summarized in Table \ref{table}. Together they sum up 
\beq
\Delta E_L=206.0243(30)
-5.2271(7)
\frac{r_p^2}{\mathrm{fm^2}}+0.0633(144)\,\mathrm{meV},\label{ELS}
\eeq 
which compared to \eq{DeltaEexp}, gives a value of the proton radius $r_p=0.8412(15)\,\rm fm$.

\nin
The first term of \eq{ELS} corresponds to the first ten entries of Table \ref{table}, which provide the QED-like contribution up to $\mathcal O (m_r \alpha^5)$, plus the leading logs at $\mathcal O (m_r \alpha^6)$ which allow us to estimate the error of this number. The main contribution is the electron vacuum polarization at $\mathcal O (m_r \alpha^3)$. The remaining amount corresponds to higher order effects such as higher loops, relativistic corrections, ultrasoft photons or perturbation theory effects. A more detailed description (with comprehensive references) of this contribution can be found in \cite{Peset:2014yha}.

\nin
The lower part of Table \ref{table} summarizes all the hadronic contributions to the Lamb shift up to the order of our interest, which we explain here in more detail. All the hadronic contributions are encoded in the $1/m^2$ potential in pNRQED:
\bea
\hspace*{-1cm}D_d^{\rm had}\!\!\! &\equiv & \!\!\!-c^{\rm had}_3-16\pi\al d^{\rm had}_2+ \frac{2\pi\al}{3}r_p^2m_p^2\,,\\
\hspace*{-1cm}\delta V_{\rm had}^{(2)}(r) \!\!\! &\equiv& \!\!\!  \frac{1}{m_p^2}D_d^{\rm had}\delta^3({\bf r})
 \rightarrow  \Delta E = -\frac{D_d^{\rm had}}{m_p^2}\frac{\left(m_r\al\right)^3}{n^3\pi }\delta_{l0}\label{Edelta}
\,.
\eea
Entries 11 and 12 correspond to the $r_p$-dependent term in \eq{ELS} (i.e. the Wilson coefficient $c_D^{\text{had}}$), and the 13th entry allows us to estimate the uncertainty of this number. The last term of \eq{ELS} comes from the two last entries of Table \ref{table}. The 14th entry of the table corresponds to the hadronic vacuum polarization (encoded in the matching coefficient $d_2^{\text{had}}$), which can be determined from dispersion relations (DR) \cite{Jegerlehner:1996ab} with a small error for our purposes, and this is the number we quote here. 

\nin
The last entry of Table \ref{table} corresponds to the two photon exchange (TPE) and deserves more care since it generates most of the uncertainty in the Lamb shift. This contribution is encoded in the Wilson coefficient $c_3^{\text{had}}$, which is unique from an EFT point of view, although it is customary to split it into Born and polarizability pieces so that $c_3^{\rm had}=c_{3}^{\rm Born}+c_{3}^{\rm pol}$. We have computed both of them separately, in the pure chiral limit and also including the contribution due to the $\Delta (1232)$, which could give the largest subleading contribution not only for being the closest resonance to the proton, but also because both of them are degenerate in the large-$N_c$ limit \cite{Dashen:1993ac}. When going from HBET to NRQED, we integrate out the pions and the Delta  and we can write $c_3^{\text{had}}\sim\alpha^2 \frac{m_\mu}{m_\pi}F(m_\pi/\Delta)+\mathcal{O}\left(\alpha^2\frac{m_\mu}{m_\rho}\right)$, where no counterterms are needed to compute the leading order of the contribution, as it is argued in Refs. \cite{Pineda:2004mx, Peset:2014yha}.
\begin{table}[htb]

\addtolength{\arraycolsep}{-0.055cm}
$$
\begin{array}{|l|l||c|r l|}
 \hline 
1&{\cal O} (m_r \alpha^3)& V_{\rm VP}^{(0)}&205.&\hspace{-0.25cm}00745  
\\ \hline
2&{\cal O} (m_r \alpha^4)& V_{\rm VP}^{(0)} & 1.&\hspace{-0.25cm}50795  
\\ \hline
3&{\cal O} (m_r \alpha^4)& V_{\rm VP}^{(0)} & 0.&\hspace{-0.25cm}15090 
\\ \hline
4&{\cal O} (m_r \alpha^5)& V_{\rm VP}^{(0)} & 0.&\hspace{-0.25cm}00752
\\ \hline
5&{\cal O} (m_r \alpha^5)& V^{(0)}_{\rm LbL} &    -0.&\hspace{-0.25cm}00089(2)
\\ \hline
6&{\cal O} (m_r \alpha^4\times \frac{m^2_\mu}{m^2_p})& V^{(2)}+V^{(3)} &   0.&\hspace{-0.25cm}05747
\\ \hline
7&{\cal O} (m_r \alpha^5)& V^{(2)}_{\rm soft}/{\rm ultrasoft} &   -0.&\hspace{-0.25cm}71902
\\ \hline
8&{\cal O} (m_r \alpha^5)& V^{(2)}_{\rm VP}        &  0.&\hspace{-0.25cm}01876
\\ \hline
9&{\cal O} (m_{\mu}\alpha^6\times \ln (\frac{m_{\mu}}{m_{e}}))& V^{(2)}; c^{(\mu)}_D &  -0.&\hspace{-0.25cm}00127
\\ \hline 
10&{\cal O} (m_{\mu}\alpha^6\times \ln \al)& V^{(2)}_{VP}; c^{(\mu)}_D&  -0.&\hspace{-0.25cm}00454
\\[0.05cm] \hline \hline 
11&{\cal O} (m_r \alpha^4\times m^2_r r^2_p)& V^{(2)}; c^{(p)}_D; r^2_p & -5.&\hspace{-0.25cm}1975
\frac{r_p^2}{\rm fm^2}
\\[0.05cm] \hline
12&{\cal O} (m_r \alpha^5\times m^2_r r^2_p)& V_{\rm VP}^{(2)}; c^{(p)}_D; r^2_p & -0.& \hspace{-0.25cm}0283
\frac{r_p^2}{\rm fm^2}  
\\[0.05cm] \hline
13&{\cal O} (m_r \alpha^6\ln\al\times m^2_r r^2_p)& V^{(2)}; c^{(p)}_D; r^2_p & -0.&\hspace{-0.25cm}0014
\frac{r_p^2}{\rm fm^2}  
\\[0.05cm] \hline
14&{\cal O} (m_r \alpha^5\times \frac{ m_r^2}{m_{\rho}^2})& V_{\rm VP_{\rm had}}^{(2)}; d^{\rm had}_2 &  0.&\hspace{-0.25cm}0111(2)
\\ \hline
15&{\cal O} (m_r \alpha^5\times \frac{m^2_r}{m^2_{\rho}}\frac{m_{\mu}}{m_{\pi}})& V^{(2)}; c^{\rm had}_3; \langle r^3 \rangle &  0.&\hspace{-0.25cm}0344(125) 
\\ \hline

\end{array}
$$
\caption{{\it The different contributions to the $\mu p$ Lamb shift in meV units.}}
\label{table}
\end{table}

\nin 
The Born contribution at leading order in the NR expansion (which guarantees that only the low energy modes contribute to the integral) reads 
\beq
\label{c3Zemach}
c_{3,\rm Born}^{pl_i}= 
4(4\pi\alpha)^2M_p^2m_{l_i}\int \frac{d^{D-1}q}{(2\pi )^{D-1}}
\frac{1 }{ {\bf q}^6}G_E^{(0)}G_E^{(2)}(-{\bf q}^2)
\,,
\eeq
where $G_E^{(0)}=1$ and $G_E^{(2)}(q^2)$ together with an analytic expression for $c_{3,\rm Born}^{pl_i}$ can be found in \cite{Peset:2014jxa}. This coefficient can also be related with (one of) the Zemach moments:
\beq
\hspace{-3.5cm}c_{3,\rm Born}^{pl_i}=\frac{\pi}{3}\al^2M_p^2m_{l_i}\langle r^3 \rangle_{(2)}
\,,\eeq
\beq
\langle r^3 \rangle_{(2)}= \frac{48}{\pi}
\int_0^{\infty} \frac{d Q }{ Q^4}
\left(
G_E^2(-Q^2)-1+\frac{Q^2}{3}\langle r^2\rangle
\right)\,.
\eeq

\nin
The Zemach moments can be determined in a similar way as the moments of the charge distribution of the proton. We have studied some and compared them to their values obtained applying DR techniques. A set of these results is summarized in Table \ref{tab:rnZemach} (a more complete discussion on this can be found in \cite{Peset:2014jxa}). 

\begin{table}[h]
$$
\begin{array}{|l||c|c|c|c|c|c|}
\hline
  & \langle r^3 \rangle  & \langle r^5 \rangle  & \langle r^7 \rangle &  \langle r^3 \rangle_{(2)}
\\ \hline\hline
\pi & 
0.4980
& 
1.619
& 
20.92
& 
0.9960
\\
\pi\&\Delta & 
0.4071
& 
1.522
& 
20.22
& 
0.8142
\\ \hline
\text{\cite{Janssens:1965kd}}& 0.7706  & 1.775   &7.006  & 2.023 \\
\text{\cite{Kelly:2004hm}}  & 
0.9838
& 
3.209
 & 
19.69
& 
2.526
\\
\text{\cite{Distler:2010zq}}  & 1.16(4) &  8.0 (1.2)(1.0) & ---& 2.85(8) \\
\hline
\end{array}
$$
\caption{{\it Values of $\langle r^n \rangle$ in fermi units. The first two rows give the prediction from the EFT at LO and LO+NLO. The third row 
in the standard dipole fit. 
The last two rows are different DR analyses.}}
\label{tab:rnZemach}
\end{table}
\nin
One would expect the chiral prediction to 
give the dominant contribution of $\langle r^n \rangle$ for $n \geq 3$ and the leading chiral log for $n=2$. Nevertheless we observe large differences (bigger than the errors) with different determinations fitting experimental data to different functions \cite{Janssens:1965kd,Kelly:2004hm,Distler:2010zq}. In this respect, the chiral result could help shaping the appropriate fit function and thus, resolving the differences between the fitted results as well as assessing their uncertainties. This difference in the fit functions has an impact on the determination of the proton radius, as can be clearly seen in Ref. \cite{Bernauer:2010wm} v.s. Refs. \cite{Lorenz:2012tm,Lorenz:2014vha} for direct fits to the $ep$ scattering data, where the determination differs in about 3-$\sigma$. In any case, the reason for such large discrepancies should be further investigated. Note that for all $n \geq 3$,  
the chiral expressions give the leading (non-analytic) dependence 
in the light quark mass as well as in $1/N_c$. This is a valuable information for eventual 
lattice simulations of these quantities where one can tune 
these parameters.

\nin
We can extract the contribution of the Born term to the energy shift from \eq{Edelta}, and this is what we quote in the last two entries of Table \ref{Table:EnZemach}. The first two entries correspond to two different DR-analyses. Note that in the HBET computation the addition of the Delta has a good convergence. On the other hand, our result is much smaller than the standard ones obtained from DR. Whether this discrepancy is due to relativistic corrections or to a need for refining the fitting procedure should be further investigated.

\begin{figure}[hbt] 
\centerline{\includegraphics[width=5.cm]{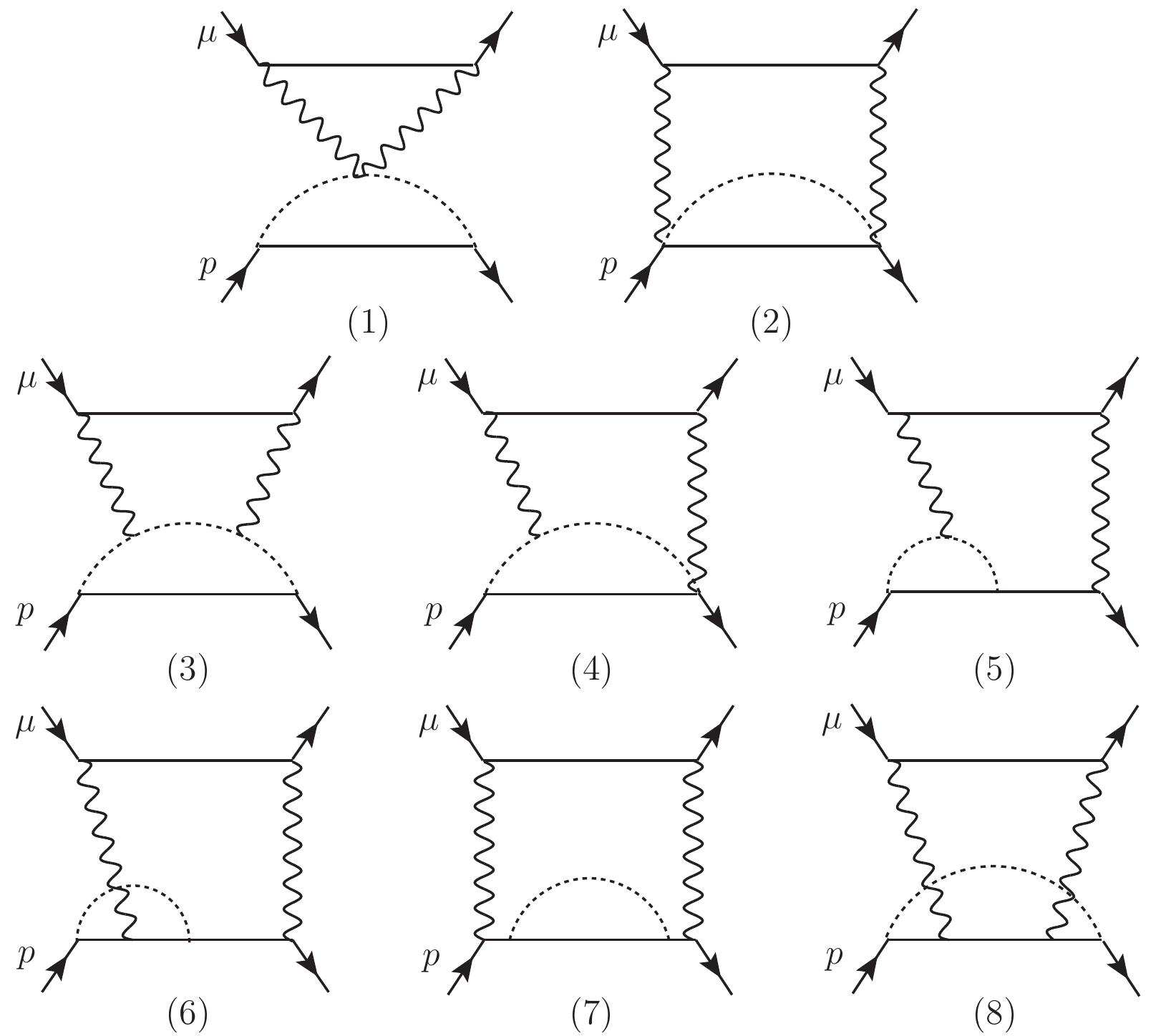}}
\caption{{\it Diagrams corresponding to the pure chiral contribution (only pions) of the TPE.} }
\label{fig1} 
\end{figure}

\begin{figure}[hbt] 
\centerline{\includegraphics[width=1.4 cm]{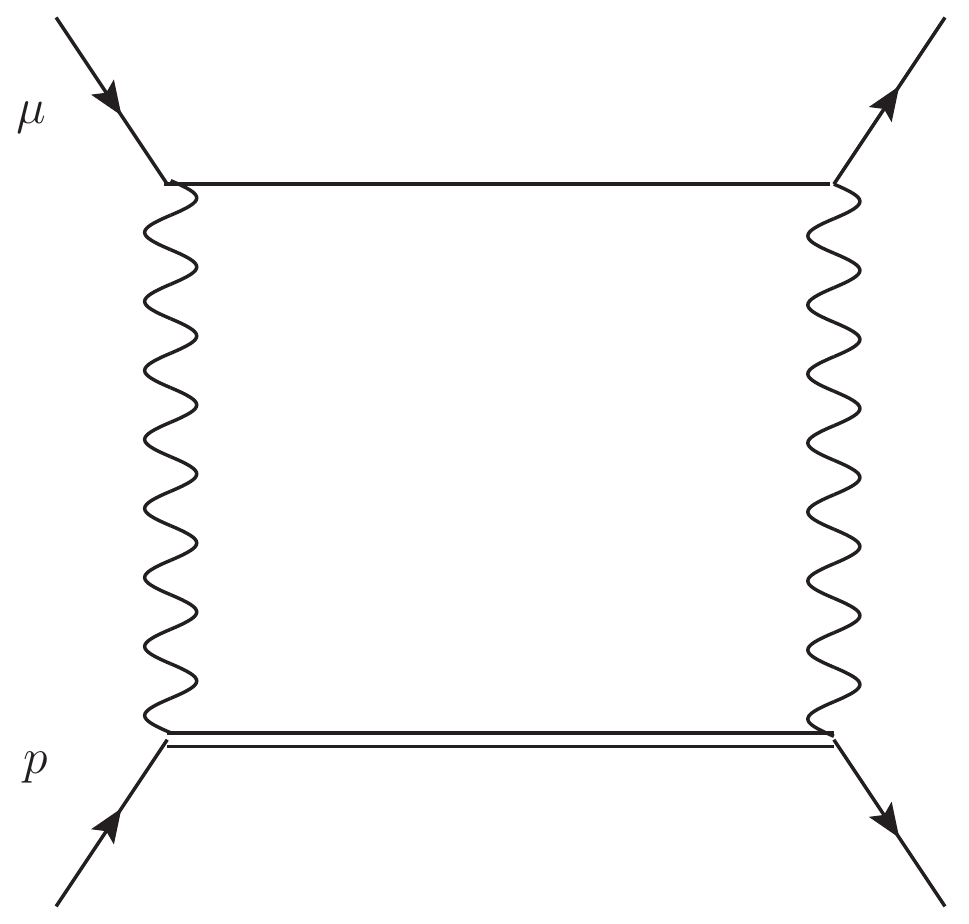}}
\centerline{\includegraphics[width=5.cm]{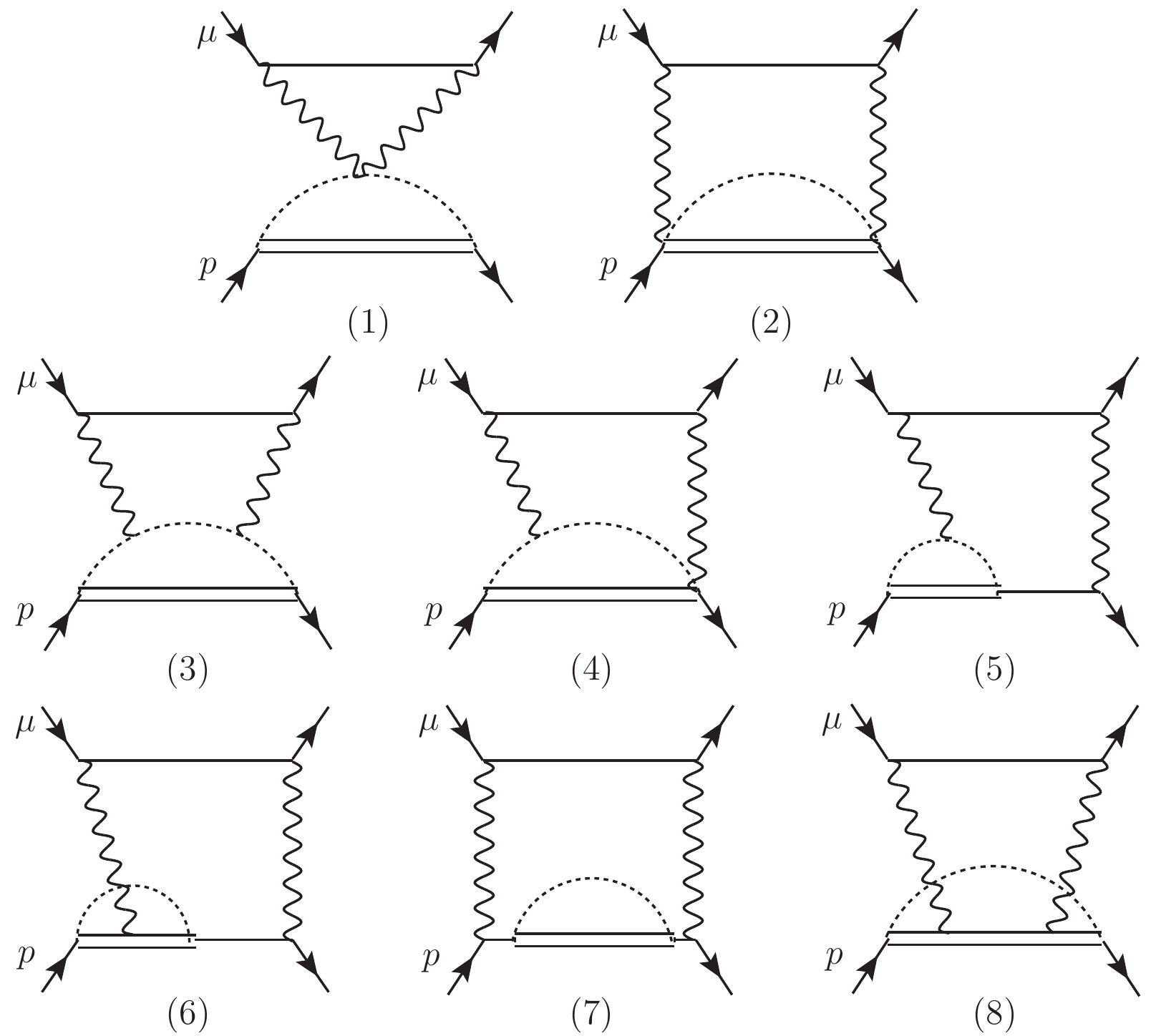}}
\caption{{\it Diagrams corresponding to the Delta tree level and loop contribution (pions \& Deltas) of the TPE.} }
\label{fig3} 
\end{figure}

\nin 
The polarizability contribution is computed through the diagrams represented in \fig{fig1} for the pure chiral case and in \fig{fig3} both for the tree level Delta exchange (top diagram) and for the one-loop Delta contribution. These diagrams are summed up in the polarizability tensor:
\beq \label{inv-dec}
\hspace{-4cm} T^{\mu\nu}_{\rm pol} =\left( -g^{\mu\nu} + \frac{q^\mu q^\nu}{q^2}\right) S_1(\rho,q^2) \nn
\eeq
\beq
  + \frac1{M_p^2} \left( p^\mu - \frac{M_p\rho}{q^2} q^\mu \right)
    \left( p^\nu - \frac{M_p\rho}{q^2} q^\nu \right) S_2(\rho,q^2)\,.
\eeq
\nin
\begin{table}[h]
\centering
$$
\begin{array}{|c|cc|cc|}                          
\hline
\mu {\rm eV}    &   \text{\cite{Pachucki:1999zza}}   &  \text{\cite{Carlson:2011zd}}   & \text{\cite{Pineda:2004mx}} (\pi) & \text{\cite{Peset:2014jxa}}(\pi\&\Delta)
\\
\hline
\Delta E_{\mathrm{Born}}   & 23.2(1.0)     &    24.7(1.6)    & 10.1(5.1) & 8.3(4.3)                                                                                          \\                                                                                            
\hline
\end{array}
$$
\caption{\it Predictions for the Born contribution to the $n=2$ Lamb shift. The first two entries correspond to DR analyses. The 
last two entries are the predictions of HBET: at LO and at LO+NLO. 
\label{Table:EnZemach}}
\end{table}

\nin
The polarizability energy shift cannot be fully obtained from DR and thus, needs some subtractions. This fact makes our model independent computation even more relevant. The Lamb shift obtained in HBET is:
\beq
\label{Epol}
\Delta E_{\mathrm{pol}}=\frac{c_{3,\rm pol}^{pl_{\mu}}}{M_p^2}\frac{1}{\pi}\left(\frac{m_r\al}{2}\right)^3=18.51(\pi)-1.58(\Delta)+9.25(\pi\Delta)\nn
\eeq
\beq
\hspace{1.5cm}=26.2(10.0)\, \mu{\rm eV}
\eeq

\nin

\begin{table}[h]
\centering
$$
\begin{array}{|c|ccc|}                          
\hline
(\mu {\rm eV})  & \text{\cite{Pachucki:1999zza}}    & \text{\cite{Carlson:2011zd}}  &  \text{\cite{Gorchtein:2013yga}}  \\
\hline
\Delta E_{\mathrm{pol}} &   12(2)          &   7.4  (2.4)   &   15.3(5.6)   
\\
\hline
\hline
(\mu {\rm eV})  & {\rm B}\chi{\rm PT} \text{\cite{Alarcon:2013cba}}  (\pi) & {\rm HBET} \text{\cite{Nevado:2007dd}} (\pi) & \text{\cite{Peset:2014yha}}(\pi\&\Delta)\\
\hline
\Delta E_{\mathrm{pol}}   & 8.2 (^{+1.2}_{-2.5}) & 18.5(9.3) & 26.2(10.0)
\\
\hline
\end{array}
$$
\caption{\it Predictions for the polarizability contribution to the $n=2$ Lamb shift. The first 3 entries use DR for the inelastic term and different modeling
functions for the subtraction term.
\label{Table:Epol}}
\end{table}
\nin
In Table \ref{Table:Epol}, we compare our HBET results to others obtained by a combination of DR for the inelastic term and different modelling functions for the subtraction term, and also to the result obtained using B$\chi$PT. This last one is carried out within a theory that treats the baryon relativistically. The result incorporates some subleading effects, which are sometimes
used to give an estimate of higher order effects in HB$\chi$PT, but it
also assumes that a theory with only baryons and pions is appropriate at the proton mass
scale, which should be taken with due caution. Still, it would be desirable to have a deeper
theoretical understanding of this difference, which may signal that relativistic corrections
are important for the polarizability correction. In any case, the B$\chi$PT computation differs
from our chiral result by around 50\% which we consider
reasonable. 

\nin
For the total TPE energy shift we obtain:
\beq
\Delta E_{\rm TPE}=\Delta E_{\rm Born}+\Delta E_{\rm pol}
=
28.59(\pi)+5.86(\pi\&\Delta)\nn\eeq \beq\hspace{3cm}=34.4(12.5)\, \mu {\rm eV}
\,,
\eeq
which, however is in good agreement with the total result \cite{Birse:2012eb} used for the determination of the proton radius in \cite{Antognini:1900ns}. This result is a pure prediction of the EFT, and it is also the most precise result that can be obtained in a model independent way since $\mathcal O (m_\mu \alpha^5\frac{m_\mu^3}{\Lambda_{\rm QCD}})$ effects are not controlled by the chiral theory and would require new counterterms.
\section{Conclusions}
\nin
We have computed in a completely model independent way the Lamb shift for $n=2$ in muonic hydrogen, which allows for the extraction of the proton radius, focusing on the hadronic contributions (mainly the TPE). Our result of the proton radius is $6.8\sigma$ away from the CODATA value and has much larger uncertainties. We have computed the pure chiral contribution to the TPE, and also the contribution due to the $\Delta(1232)$. This computation of the TPE gives a similar result to the one obtained by the combination of DR plus the use of different models. However, the partial computations (Born and polarizability) differ from the partial results obtained in these frameworks, fact that should be further understood.
\section*{Acknowledgements}
\nin 
The author thanks Antonio Pineda for his collaboration in the development of this work. This work was supported by the Spanish grant  FPA2011-25948 and the Catalan grant SGR2009-00894.






\begin{thebibliography}{00}


\bibitem{Pohl:2010zza}
  R.~Pohl {\it et al.},
  Nature {\bf 466} (2010) 213.

\bibitem{Antognini:1900ns} 
  A.~Antognini, F.~Nez, K.~Schuhmann, F.~D.~Amaro, FrancoisBiraben, J.~M.~R.~Cardoso, D.~S.~Covita and A.~Dax {\it et al.},
  Science {\bf 339}, 417 (2013).

\bibitem{Mohr:2012tt} 
  P.~J.~Mohr, B.~N.~Taylor and D.~B.~Newell,
  Rev.\ Mod.\ Phys.\  {\bf 84}, 1527 (2012)
  [arXiv:1203.5425 [physics.atom-ph]].

\bibitem{Lorenz:2014vha}
  I.~T.~Lorenz, H.-W.~Hammer and U.~G.~Meissner,
  Eur.\ Phys.\ J.\ A {\bf 48} (2012) 151
  [arXiv:1205.6628 [hep-ph]].
  
  I.~T.~Lorenz and U.-G.~Meissner,
  Physics Letters B (2014), pp. 57-59
  [arXiv:1406.2962 [hep-ph]].

  
\bibitem{Jenkins:1990jv} 
  E.~E.~Jenkins and A.~V.~Manohar,
  Phys.\ Lett.\ B {\bf 255}, 558 (1991).
  
\bibitem{Bernard:1992qa}
  V.~Bernard, N.~Kaiser, J.~Kambor and U.~G.~Meissner,
  Nucl.\ Phys.\ B {\bf 388} (1992) 315.

\bibitem{Caswell:1985ui} 
  W.~E.~Caswell and G.~P.~Lepage,
  Phys.\ Lett.\ B {\bf 167}, 437 (1986).

\bibitem{Pineda:1997bj}
  A.~Pineda and J.~Soto,
  Nucl.\ Phys.\ Proc.\ Suppl.\  {\bf 64}, 428 (1998)
  [arXiv:hep-ph/9707481].

\bibitem{Pineda:1997ie} 
  A.~Pineda and J.~Soto,
  Phys.\ Lett.\ B {\bf 420}, 391 (1998)
  [hep-ph/9711292].

\bibitem{Pineda:1998kn}
  A.~Pineda and J.~Soto,
  Phys.\ Rev.\  D {\bf 59}, 016005 (1999)
  [arXiv:hep-ph/9805424].
  
\bibitem{Pineda:2002as}
  A.~Pineda,
  Phys.\ Rev.\  C {\bf 67}, 025201 (2003)
  [arXiv:hep-ph/0210210];
  A.~Pineda,
  hep-ph/0308193.
 
\bibitem{Pineda:2004mx}
  A.~Pineda,
  Phys.\ Rev.\  C {\bf 71}, 065205 (2005)
  [arXiv:hep-ph/0412142].
  
\bibitem{Nevado:2007dd}
  D.~Nevado and A.~Pineda,
  Phys.\ Rev.\  C {\bf 77}, 035202 (2008)
  [arXiv:0712.1294 [hep-ph]].

\bibitem{Pineda:2011xp} 
  A.~Pineda,
  arXiv:1108.1263 [hep-ph].


\bibitem{Peset:2014yha}
  C.~Peset and A.~Pineda,
  arXiv:1403.3408 [hep-ph].
\bibitem{Peset:2014jxa}
  C.~Peset and A.~Pineda,
  Nucl.\ Phys.\ B  {\bf 887}, 69 (2014)
 [arXiv:1406.4524 [hep-ph]].

\bibitem{Jegerlehner:1996ab} 
  F.~Jegerlehner,
  Nucl.\ Phys.\ Proc.\ Suppl.\  {\bf 51C}, 131 (1996)
  [hep-ph/9606484].
  
\bibitem{Dashen:1993ac} 
  R.~F.~Dashen and A.~V.~Manohar,
  Phys.\ Lett.\ B {\bf 315}, 438 (1993)
  [hep-ph/9307242].
  
\bibitem{Janssens:1965kd} 
  T.~Janssens, R.~Hofstadter, E.~B.~Hughes and M.~R.~Yearian,
  Phys.\ Rev.\  {\bf 142}, 922 (1966).

\bibitem{Kelly:2004hm} 
  J.~J.~Kelly,
  Phys.\ Rev.\ C {\bf 70}, 068202 (2004).
  
\bibitem{Distler:2010zq} 
  M.~O.~Distler, J.~C.~Bernauer and T.~Walcher,
  Phys.\ Lett.\ B {\bf 696}, 343 (2011)
  [arXiv:1011.1861 [nucl-th]].
  
\bibitem{Pachucki:1999zza} 
  K.~Pachucki,
  Phys.\ Rev.\ A {\bf 60}, 3593 (1999).
  

\bibitem{Carlson:2011zd} 
  C.~E.~Carlson and M.~Vanderhaeghen,
  Phys.\ Rev.\ A {\bf 84}, 020102 (2011)
  [arXiv:1101.5965 [hep-ph]].
  
  
\bibitem{Gorchtein:2013yga} 
  M.~Gorchtein, F.~J.~Llanes-Estrada and A.~P.~Szczepaniak,
  Phys.\ Rev.\ A {\bf 87}, 052501 (2013)
  [arXiv:1302.2807 [nucl-th]].
  
\bibitem{Alarcon:2013cba}
  J.~M.~Alarcon, V.~Lensky and V.~Pascalutsa,
  Eur.\ Phys.\ J.\ C {\bf 74} (2014) 2852
  [arXiv:1312.1219 [hep-ph]].
  
\bibitem{Bernauer:2010wm} 
  J.~C.~Bernauer {\it et al.}  [A1 Collaboration],
  Phys.\ Rev.\ Lett.\  {\bf 105}, 242001 (2010)
  [arXiv:1007.5076 [nucl-ex]].
  
\bibitem{Lorenz:2012tm} 
  I.~T.~Lorenz, H.~-W.~Hammer and U.~-G.~Meissner,
  Eur.\ Phys.\ J.\ A {\bf 48}, 151 (2012)
  [arXiv:1205.6628 [hep-ph]].
  
  \bibitem{Birse:2012eb} 
 M.~C.~Birse and J.~A.~McGovern,
Eur.\ Phys.\ J.\ A {\bf 48}, 120 (2012)
  [arXiv:1206.3030 [hep-ph]].
  
  
  





 


 


 









 \end{thebibliography}



\end{document}